\documentclass[twocolumn]{aastex63}

\usepackage{amssymb, amsfonts, amsmath,framed}
\usepackage{comment}
\usepackage{enumitem}
\usepackage{comment}

\usepackage{bm}
\expandafter\ifx\csname package@font\endcsname\relax\else
 \expandafter\expandafter
 \expandafter\usepackage
 \expandafter\expandafter
 \expandafter{\csname package@font\endcsname}
\fi
\expandafter\ifx\csname package@font\endcsname\relax\else
 \expandafter\expandafter
 \expandafter\usepackage
 \expandafter\expandafter
 \expandafter{\csname package@font\endcsname}
\fi
\def\bi{\begin{itemize}}
\def\ei{\end{itemize}}
\def\bq{\begin{equation}}
\def\eq{\end{equation}}
\def\bqy{\begin{eqnarray}}
\def\eqy{\end{eqnarray}}

\begin{document}
\title{Ultra-High-Energy Cosmic Rays Accelerated by Magnetically Dominated Turbulence}

\author{Luca Comisso}
\affiliation{Department of Physics, Columbia University, New York, NY 10027, USA}
\affiliation{Department of Astronomy, Columbia University, New York, NY 10027, USA}
\affiliation{Columbia Astrophysics Laboratory, Columbia University, New York, NY 10027, USA}

\author{Glennys R. Farrar}
\affiliation{Center for Cosmology and Particle Physics, Department of Physics, New York University, New York, NY 10003, USA}

\author{Marco S. Muzio}
\affiliation{Department of Physics, Pennsylvania State University, University Park, PA 16802, USA}
\affiliation{Department of Astronomy and Astrophysics, Pennsylvania State University, University Park, PA 16802, USA}
\affiliation{Institute of Gravitation and the Cosmos, Center for Multi-Messenger Astrophysics, Pennsylvania State University, University Park, PA 16802, USA}

\begin{abstract}

Ultra-High-Energy Cosmic Rays (UHECRs), particles characterized by energies exceeding $10^{18}$ eV, are generally believed to be accelerated electromagnetically in high-energy astrophysical sources. One promising mechanism of UHECR acceleration is magnetized turbulence. We demonstrate from first principles, using fully kinetic particle-in-cell simulations, that magnetically dominated turbulence accelerates particles on a short timescale, producing a power-law energy distribution with a rigidity-dependent, sharply defined cutoff well approximated by the form $f_{\rm cut}\left({E, E_{\rm cut}}\right) = {\text{sech}}\left[ ( {{E}/{E_{\rm cut}}} )^2 \right]$. Particle escape from the turbulent accelerating region is energy-dependent, with $t_{\rm esc} \propto E^{-\delta}$ and $\delta \sim 1/3$. The resulting particle flux from the accelerator follows $dN/dEdt \propto E^{-s} {\text{sech}}\left[ ( {{E}/{E_{\rm cut}}} )^2 \right]$, with $s \sim 2.1$. We fit the Pierre Auger Observatory's spectrum and composition measurements, taking into account particle interactions between acceleration and detection, and show that the turbulence-associated energy cutoff is well supported by the data, with the best-fitting spectral index being $s = 2.1^{+0.06}_{-0.13}$. Our first-principles results indicate that particle acceleration by magnetically dominated turbulence may constitute the physical mechanism responsible for UHECR acceleration.

\vspace{0.9cm}

\end{abstract}

\section{Introduction} \label{sec:intro} 

The search for the sources of Ultra-High-Energy Cosmic Rays (UHECRs) is a central pursuit in astroparticle physics. While the bulk of Galactic cosmic rays are thought to be accelerated by the forward shocks of supernova remnants \citep{Hillas05,Blasi13}, the origin and acceleration mechanism of the highest-energy particles remain elusive and are actively debated \citep{Coleman2023}.  
Over the past decade, experimental advances have allowed the features of the all-particle energy spectrum and nuclear composition to be studied with unprecedented precision \citep{Aab+20b,TA_2023APh,TA_2024PhRvD,PA_2024arXiv}, laying the groundwork for the development of theoretical models seeking to explain these observations.

The all-particle cosmic-ray spectrum at the highest energies reveals several distinctive features. The ``ankle'', observed at $\sim5\times10^{18}$ eV, is marked by a pronounced hardening of the spectrum. This is followed by the ``instep'' at $\sim1.3\times10^{19}$ eV, where the spectrum steepens. Finally, a sharp suppression is observed above $\sim5\times10^{19}$ eV. The nuclear composition and its energy evolution can be inferred from the depth of maximum development of individual air showers, $X_{\rm max}$, specifically by the mean and variance of the ensemble of showers. Below the ankle, extragalactic cosmic rays are primarily protons and helium. As the energy increases, the composition becomes progressively heavier, consistent with a particle acceleration process constrained by particle rigidity, $\mathcal{R} \simeq E/eZ$. 
At each energy, the distribution of nuclear masses in the extragalactic component is remarkably narrow, as indicated by the small dispersion of $X_{\rm max}$ values around the mean. 

Analyzing different astrophysical source scenarios and fitting the aforementioned observational data provides a means to infer the properties of the cosmic ray spectrum at the sources \citep{Aloisio2014}. 
Inspired by theoretical predictions for shock acceleration spectra \citep[e.g.][]{ProtheroeStanev99}, most studies sumed that source spectra follow a power-law energy distribution with an exponential cutoff, $\phi(E) \propto E^{-s} \exp\left(-E/{E_{\rm cut}}\right)$. With this form, fits to the data~\citep{Aab2017JCAP} returned spectra much harder (smaller $s$) than expected for the standard acceleration mechanisms, which generally predict $s \gtrsim 2$ \citep{BE87,Longair2011}.  The hard spectral indices are required in the fits to suppress the low energy tails of the heavier components that dominate at higher energies, for consistency with the observed small variance of $X_{\rm max}$ at any given $E$.  To address this conflict with theoretical expectations, several alternative cutoff functions—though not physically motivated—have been used in an ad hoc effort to assess the sensitivity of the best-fitting power-law to the cutoff \citep[e.g.][]{Aab2017JCAP,Halim2023JCAP,augerMagHoriz24}. Among them, the broken-exponential cutoff, with a pure power law below the break energy and a pure exponential above, has become a popular choice, but nonetheless yields $s \sim 1$ for the reference fits in \citet{Aab2017JCAP} and \citet{Halim2023JCAP}. In some scenarios, the fits even result in negative values for $s$ \citep[e.g.][]{Aab2017JCAP}, further complicating the theoretical understanding.

In this work, we adopt a rigorous physics-based approach, leveraging large-scale fully kinetic simulations of the plasma dynamics to determine the properties of the particle energy spectrum at the source. We demonstrate that magnetically dominated turbulence accelerates ions into power-law energy spectra with spectral index around $2$ and a sharp cutoff well-approximated by $\phi(E) \propto E^{-s}\, {\text{sech}}\left[ ( {{E}/{E_{\rm cut}}} )^2 \right]$. We further show that magnetically dominated turbulence accelerates particles on a very short timescale and efficiently converts a significant fraction of turbulent energy into relativistic ions. Finally, we compare the UHECR spectrum and composition at Earth from the Pierre Auger Observatory to predictions using the turbulence-based energy cutoff and predicted spectral index of escaping particles and find good agreement with the data. These results establish magnetically dominated turbulence as a compelling physical mechanism capable of resolving the challenges of UHECR acceleration.

\section{Fully-Kinetic Model} \label{sec:model} 

In the collisionless limit, the kinetic dynamics of a plasma is governed by the Vlasov–Maxwell system of equations. 
These equations describe the evolution of the particle distribution function $f_{\alpha}(\bm{x}, \bm{p}, t)$ for each particle species ${\alpha}$ at position $\bm{x}$ and momentum $\bm{p}$, together with the self-consistent electric field $\bm{E}(\bm{x}, t)$ and magnetic field $\bm{B}(\bm{x}, t)$. They evolve according to 
\begin{equation}
\begin{aligned}
  {\partial f_{\alpha}}/{\partial t} &= -\bm{v} \cdot \bm{\nabla}f_{\alpha} - q_{\alpha} (\bm{E} + {\bm{v} \times \bm{B}}/c) \cdot {\partial f_{\alpha}}/{\partial\bm{p}} \, , \\
  {\partial \bm{E}}/{\partial t} &= c \bm{\nabla} \times \bm{B} - 4 \pi \bm{j} \, , \\
  {\partial \bm{B}}/{\partial t} &= -c \bm{\nabla} \times \bm{E} \, , 
\end{aligned}
\end{equation}
with the constraints $\bm{\nabla} \cdot \bm{E} = 4 \pi \rho$ and $\bm{\nabla} \cdot \bm{B} = 0$. The charge density $\rho(\bm{x}, t)$ and current density $\bm{j}(\bm{x}, t)$ are given by
\begin{equation}
\begin{aligned}
  \rho &= \sum\limits_{\alpha} q_{\alpha} \int{d^3\bm{p} \, f_{\alpha}} \, , \\
  \bm{j} &= \sum\limits_{\alpha} q_{\alpha} \int{d^3\bm{p} \, \bm{v} f_{\alpha}} \, , 
\end{aligned}
\end{equation}
where $q_{\alpha}=Ze$ is the species charge, $e$ is the elementary charge, and $\bm{v}$ is the particle velocity. We solved this system along characteristics of the Vlasov equation with the Particle-in-Cell (PIC) method \citep{birdsall_langdon_85} employing the massively parallel code TRISTAN-MP \citep{buneman_93, spitkovsky_05}. The simulations do not include energy losses in the accelerator, consistent with the absence of pile-up in the observed spectrum of UHECRs~\citep{Protheroe04,zirakashviliAharonian06}. 

We performed the simulations in a triply periodic cubic domain of size $L^3$. The plasma consisted of electrons and ions, with combined particle density $n_{e0} + n_{i0} = n_0$. 
We conducted the PIC simulations with a single ion species of charge number $Z=1$ and mass $m_i = 1836 \, m_e$, i.e. protons.
Computational particles were initialized according to a Maxwell-J\"{u}ttner distribution with temperature $T_{e0} = T_{i0} = T_0$, where $T_{e0}$ and $T_{i0}$ are the electron and ion temperatures, respectively. 
A uniform mean magnetic field ${\bm{B}}_0  = B_0 {\bm{\hat e}}_z$ was imposed in the $z$ direction. 
Analogously to \citet{Comisso18,Comisso19,Comisso21}, we seeded turbulence by initializing a spectrum of magnetic fluctuations in the form $\delta {\bm{B}}({\bm{x}}) = \sum_{\bm{k}} {\delta B_{\bm{k}} {\bm{\hat \xi}}_{\bm{k}} \exp \left[ {i \left( {{\bm{k}} \cdot {\bm{x}} + {\phi_{\bm{k}}}} \right)} \right]}$, where $\delta B_{\bm{k}}$ is the Fourier amplitude of the mode with wave vector ${\bm{k}}$, ${\bm{\hat \xi}}_{\bm{k}} = i \, {\bm{k}} \times {\bm{B}}_0/|{\bm{k}} \times {\bm{B}}_0|$ are Alfv{\'e}nic polarization unit vectors, and $\phi_{\bm{k}}$ are random phases. To ensure that $ \delta {\bm{B}}({\bm{x}})$ is a real function, we set $\delta B_{\bm{k}} = \delta B_{-{\bm{k}}}$ and $\phi_{\bm{k}} = - \phi_{-{\bm{k}}}$. 
We assigned equal amplitude per mode, with wave vector components $k_j = 2\pi n_j/L$ where $j=x,y,z$ and we consider $n_j =1,2,3$.
With these choices, the initial outer coherence length scale of the magnetic field is approximately $l_c = L/3$. 

The strength of the initial magnetic field fluctuations is parameterized by the magnetization associated with the fluctuations, $\sigma_{\delta B}= {\delta B_0^2}/{4\pi h_0}$, where $\delta B_0 = \langle |{\delta {\bm{B}}({\bm{x}})}|^2 \rangle^{1/2}$ is the root-mean-square value of the fluctuations, and $h_0$ is the enthalpy density, accounting for both ion and electron contributions. We consider strong turbulence with $\delta B_0 = B_0$.  The corresponding total magnetization is $\sigma = {B^2_{\rm rms}}/{4\pi h_0} = 2 \sigma_{\delta B}$, where $B_{\rm rms} = \langle |{{\bm{B}}({\bm{x}})}|^2 \rangle^{1/2}$. Since we are interested in the magnetically dominated regime, we take $\sigma_{\delta B} > 1$,  which yields the Alfv{\'e}n speed ${v_{A}} = c [{\sigma}/(1 + {\sigma})]^{1/2} \simeq c$.  We conducted simulations with 5 different values of plasma magnetization, $\sigma \in \left\{ {8,16,32,64,128} \right\}$.  We also performed simulations with 3 different domain sizes $L/d_i \in \left\{ {250,400,640} \right\}$, where $d_i=c/\omega_{pi}$ is the ion inertial length, with $\omega_{pi} = ({4\pi n_0 {e^2}/{\gamma_{0,i} m_i}})^{1/2}$ indicating the ion plasma frequency and $\gamma_{0,i}$ being the mean ion Lorentz factor. 
We initialize the ion temperature to $k_B T_0 = 0.2 m_i c^2$, yielding $\gamma_{0,i} \simeq 1.36$. This parameter choice does not impact our findings  \citep{Comisso19}. 
We adopt uniform grid cells of size $\Delta x = \Delta y = \Delta z = d_{e0}/2$, where $d_{e}=c/\omega_{pe}$ is the initial electron inertial length, with $\omega_{pe} = (4\pi n_0 {e^2}/\gamma_{0,e} m_e)^{1/2}$ indicating the electron plasma frequency. 
We use an average of $20$ computational particles per cell, which adequately resolves the nonthermal particle acceleration process, as shown in \citet{Comisso18,Comisso19}. The time step is chosen according to the Courant-Friedrichs-Lewy condition, and all simulations are run for $t \geq 15 l_c/c$ to ensure that the particle energy spectrum has reached a steady state. We take the simulation with $\sigma=16$ and $L/d_i = 400$ as the reference simulation.

\begin{figure}
\begin{center}
    \includegraphics[width=8.55cm]{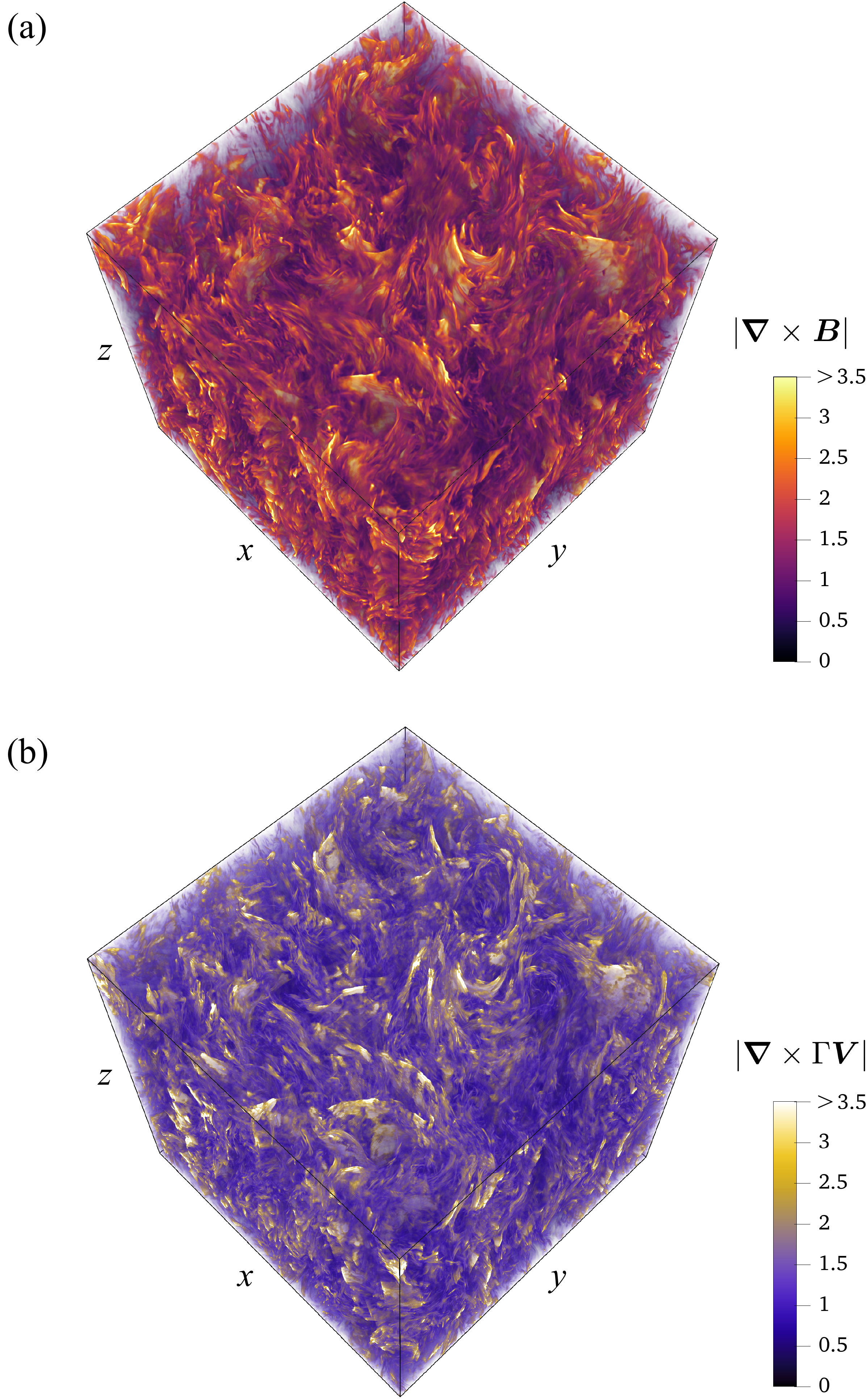}

\vspace{0.45cm}
    
    \includegraphics[width=8.65cm]{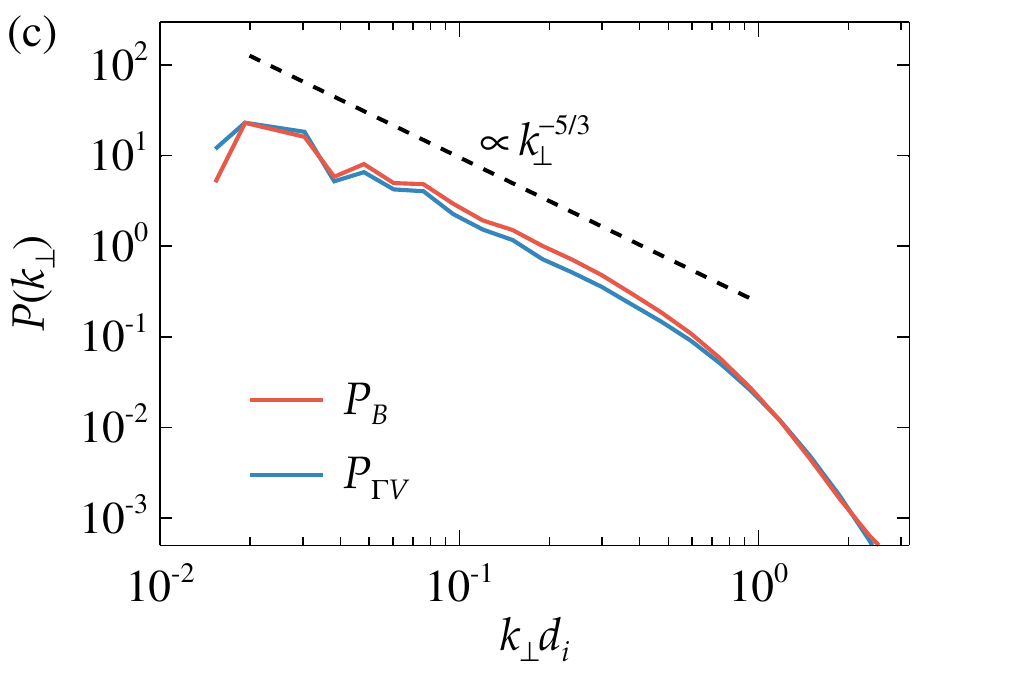}
\end{center}
\vspace{-0.3cm}
    \caption{Volume rendering of (a) $|\bm{\nabla} \times \bm{B}|$ and (b) $|\bm{\nabla} \times \Gamma \bm{V}|$ (normalized to their respective root-mean-square values) taken at $t = 2.5 l_c/c$ from the reference turbulence simulation ($\sigma = 16$, $l_c/d_i = 133$). (c) One-dimensional spectra of the turbulent magnetic field (red) and fluid four-velocity (blue) at $t = 2.5 l_c/c$. Each spectrum is normalized so that $\sum_{k_\bot}P(k_\bot) = 1$. A power-law slope of $k_\bot^{-5/3}$ (dashed black line) is shown for reference.}
\label{fig1}
\end{figure}

In our simulations, turbulence develops from the initialized magnetic fluctuations, and then decays in time as no continuous driving is imposed. A well-developed turbulent cascade is established on the outer-scale nonlinear timescale $l_c/c$. In Figures \ref{fig1}(a) and \ref{fig1}(b), we show a volume rendering of the vortical fields $|\bm{\nabla} \times \bm{B}|$ and $|\bm{\nabla} \times \Gamma \bm{V}|$, taken at time $t = 2.5 l_c/c$ from the reference simulation. Here, $\Gamma = [1-(\bm{V}/c)^2]^{-1/2}$ is the plasma fluid Lorentz factor and $\bm{V}$ is the plasma fluid velocity obtained by averaging the velocities of individual particles. Sheet-like structures are ubiquitous in the turbulent domain for both fields. Due to the presence of the mean field ${\bm{B}}_0  = B_0 {\bm{\hat e}}_z$, these structures are mostly elongated along ${\bm{\hat e}}_z$. Our large-scale PIC simulations capture the turbulent cascade from MHD scales to kinetic scales. In Fig.~\ref{fig1}(c), we present the one-dimensional spectra $P_X(k_\bot) dk_\bot = \sum_{{\bm{k}} \in dk_\bot} {{\bm{X}_{\bm{k}} \cdot {\bm{X}}_{\bm{k}}^*}}$ for the magnetic field fluctuations (${\bm{X}} = \delta {\bm{B}}$) and the fluid four-velocity fluctuations (${\bm{X}} = \Gamma \bm{\bm{V}}$). Here $k_\bot = {(k_x^2 + k_y^2)^{1/2}}$ represents wavenumbers perpendicular to ${\bm{B}}_0$. The turbulent cascade exhibits an extended MHD range, with $P_{B,\Gamma V}(k_\bot) \propto k_\bot^{-5/3}$ for $k_\bot d_{i} \lesssim 1$, consistent with turbulence scaling theories \citep{GS95,TB98}. At scales smaller than $d_{i}$, both spectra steepen as the turbulent cascade progresses into the kinetic range \citep[e.g.][]{Chen14}.

\section{Nonthermal Particle Acceleration} 

Previous large-scale PIC simulations have demonstrated that large-amplitude turbulence ($\delta B_0 \sim B_0$) in a highly magnetized plasma ($\sigma \gg 1$) produces a significant population of nonthermal particles with a power-law energy distribution that is independent of microscopic kinetic scales \citep{Comisso18, Comisso19} (for results on moderately magnetized turbulence, see \citet{Zhd18}). Here, we focus on the particle energy cutoff to determine whether magnetized turbulence can serve as the mechanism responsible for UHECR acceleration, as the functional shape of the energy  cutoff is closely tied to the UHECR spectrum and composition observables.

Particle acceleration via magnetized turbulence is affected by the particle's magnetic rigidity ${\mathcal{R}} = |{\bm{p}}| c/q$, 
as the interaction with turbulence fluctuations depends on the Larmor radius $\rho_L \simeq |{\bm{p}}| c/qB$ (taking $|{\bm{p}}| \simeq p_\bot$). Particles are effectively accelerated until their Larmor radius become comparable to the coherence length of the magnetic field, i.e., $\rho_L ({\mathcal{R}}_{\rm cut}) = \kappa l_c$, with $\kappa = \mathcal{O}(1)$. For ultra-relativistic particles, this translates to a cutoff energy estimate 
\begin{equation} \label{eq:Ecut} 
  E_{\rm cut} = Z e {\mathcal{R}}_{\rm cut} = Z e \,B_{\rm rms}\, \kappa \,l_c \, .
\end{equation} 
In the following, we show that $\kappa = \mathcal{O}(1)$ and that the particle energy spectrum is characterized by a very sharp cutoff at this energy.

\begin{figure}
\begin{center}
    \includegraphics[width=8.65cm]{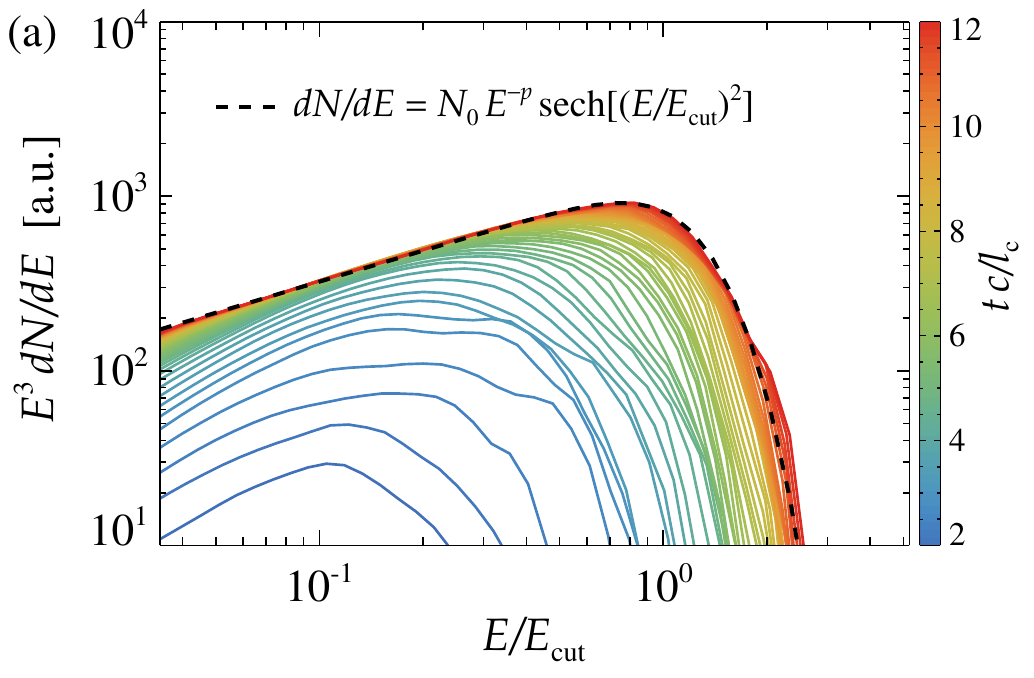}

\vspace{0.2cm}
    
    \includegraphics[width=8.65cm]{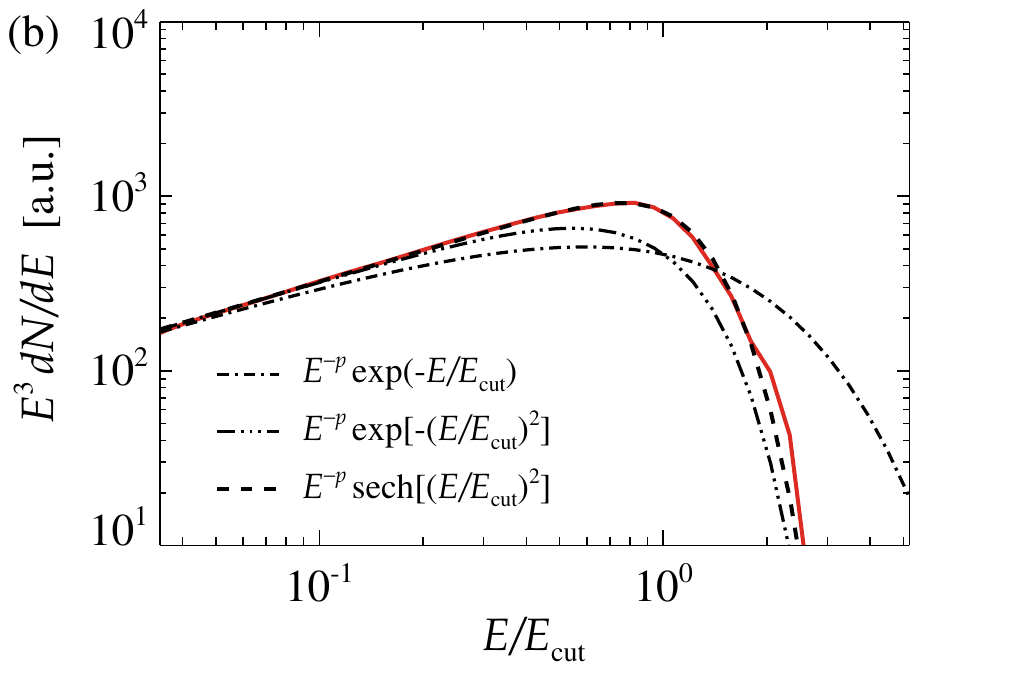}
\end{center}
\vspace{-0.4cm}
    \caption{(a) Time evolution of the ion energy spectrum $dN/dE$, multiplied by $E^3$, for the reference turbulence simulation ($\sigma = 16$, $l_c/d_i = 133$). At late times, when the particle energy spectrum ceases to evolve, it exhibits a power-law distribution with a sharp cutoff. The spectrum is well-fitted by the function $N_0 E^{-p} {\text{sech}}\left[ ( {{E}/{E_{\rm cut}}} )^2 \right]$, where $p = 2.4$, $N_0$ is a normalization factor, and $E_{\rm cut}$ is the cutoff energy given by Eq. (\ref{eq:Ecut}), with $\kappa=0.65$. (b) $E^3 dN/dE$ at saturation, $t = 12 l_c/c$ (solid red line), overlaid with the functional forms $E^{-p} \exp\left(-E/{E_{\rm cut}}\right)$ (dotted-dashed black line), $E^{-p} \exp \left[-({{E}/{E_{\rm cut}}} \right)^2]$ (triple-dotted–dashed black line), and $E^{-p} {\text{sech}}\left[ ( {{E}/{E_{\rm cut}}} )^2 \right]$ (dashed black line).}
\label{fig2}
\end{figure}

In Figure \ref{fig2}(a), we show the time evolution of the
ion energy spectrum $dN/dE$, multiplied by $E^3$, for the reference turbulence simulation. At late times, when the turbulent energy has predominantly transferred to the particles, the spectrum ceases evolving (orange and red lines) and can be described by a relation of the form 
\begin{equation} \label{eq:general} 
  \frac{dN}{dE} = N_0 E^{-p} f_{\rm cut}\left({E, E_{\rm cut}}\right)  \, ,
\end{equation} 
with a cutoff that closely follows
\begin{equation} \label{eq:fcut} 
  f_{\rm cut}\left({E, E_{\rm cut}}\right) = {\text{sech}}\left[ \left( {\frac{E}{E_{\rm cut}}} \right)^2 \right] \, ,
\end{equation} 
with $\kappa=0.65$ in the fit. 
In Figure \ref{fig2}(b), we compare the fit provided by the cutoff given by Eq. (\ref{eq:fcut}) with the simple exponential cutoff $f_{\rm cut} = \exp\left(-E/{E_{\rm cut}}\right)$. We also show $f_{\rm cut} = \exp \left[-({{E}/{E_{\rm cut}}} \right)^2]$, which approximates the behavior of Eq. (\ref{eq:fcut}) for $E \gg E_{\rm cut}$. Both functions exhibit smoother behavior compared to the actual cutoff observed in the simulation, with the exponential cutoff $\exp\left(-E/{E_{\rm cut}}\right)$ significantly failing to represent the PIC simulation results.

\begin{figure}
\begin{center}
   \includegraphics[width=8.65cm]{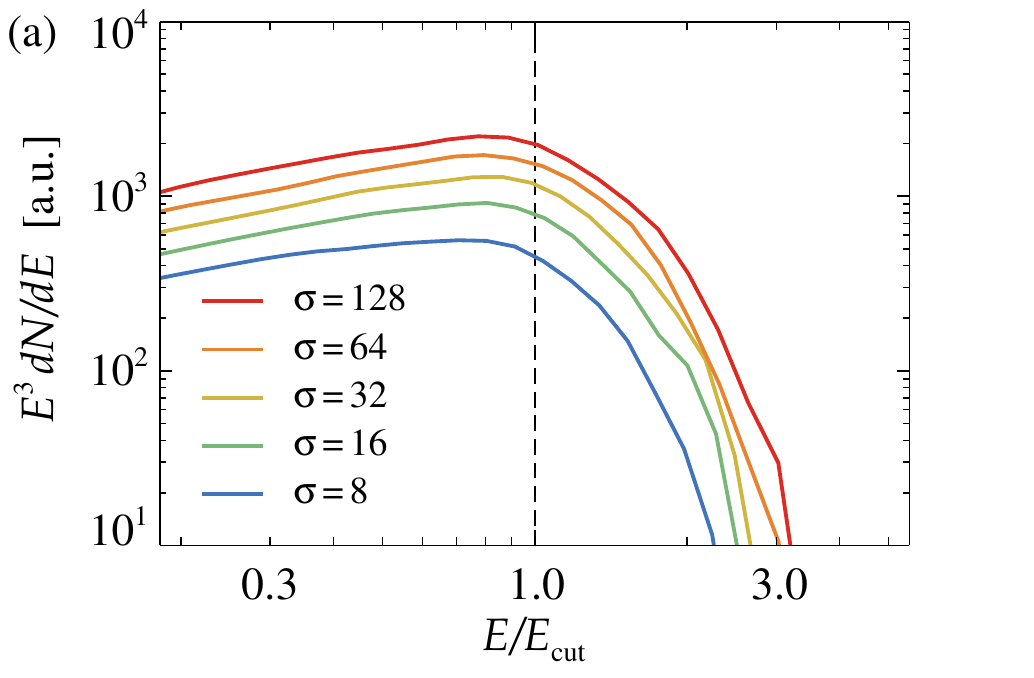}

\vspace{0.15cm}
    
    \includegraphics[width=8.65cm]{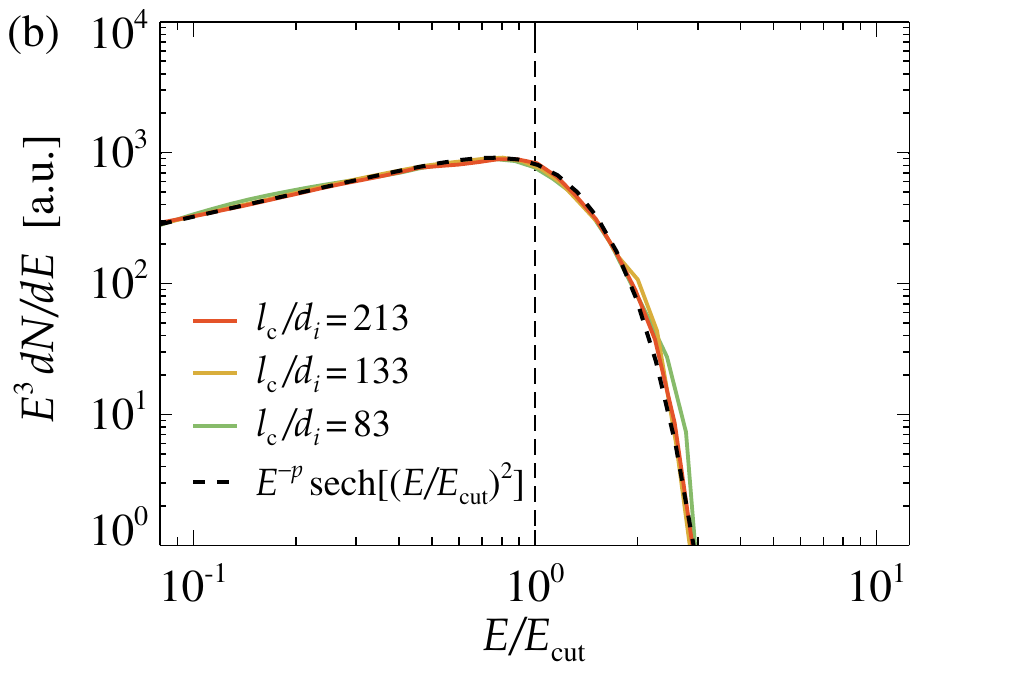}
    
\vspace{0.15cm}
    
    \includegraphics[width=8.65cm]{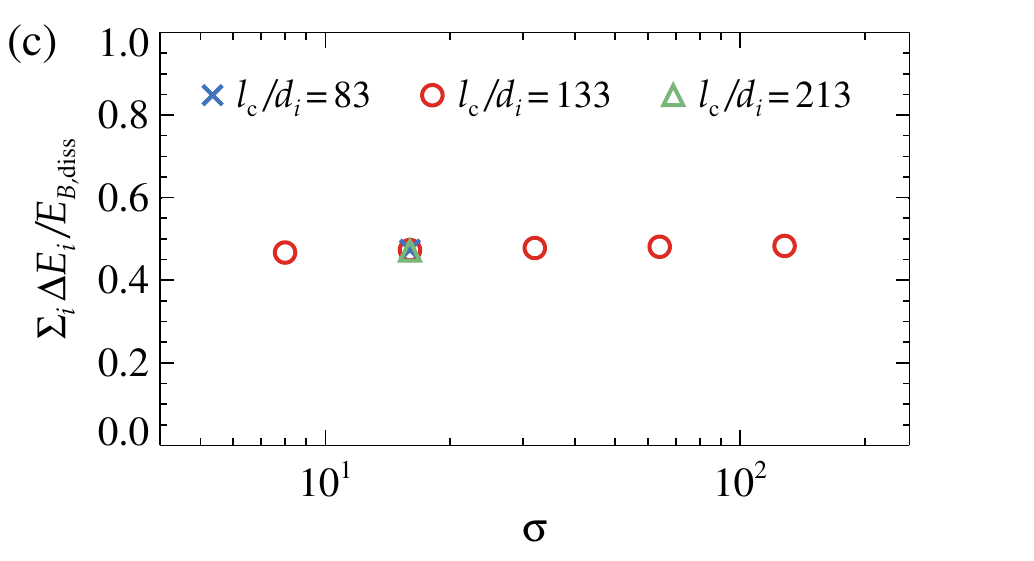}
    
\end{center}
\vspace{-0.5cm}
    \caption{(a) Ion energy spectra at saturation ($t = 12 l_c/c$) for turbulence simulations with varying magnetization $\sigma = (8, 16, 32, 64, 128)$ and fixed ${l_c}/d_i = 133$. In all cases, the particle energy spectrum exhibits a cutoff at $E_{\rm cut}$ as given by Eq. (\ref{eq:Ecut}). (b) $E^3 dN/dE$ for simulations with varying $l_c/d_i = (83, 133, 213)$ and fixed $\sigma = 16$, overlaid with $dN/dE \propto E^{-p} {\text{sech}}\left[ ( {{E}/{E_{\rm cut}}} )^2 \right]$, where $p=2.4$ (dashed black line). (c) Fraction of dissipated magnetic energy, $E_{B,{\rm diss}}$, converted to ion energy at $t=12 l_c/c$ across the different simulations in our study.}
\label{fig3}
\end{figure}

In Figure \ref{fig3}, we compare the late-time particle energy spectra from simulations with varying magnetization $\sigma$ (frame (a)) and different turbulence inertial ranges $l_c/d_i$ (frame (b)). In these simulations, $\sigma$ is varied by changing the magnetic field strength ($\sigma \propto B^2_{\rm rms}$), while $l_c/d_i$ is varied by changing $l_c$. Figs. \ref{fig3}(a) and \ref{fig3}(b) show that all simulations produce the same sharp cutoff. When the cutoff energy $E_{\rm cut}$ is accounted for using Eq. (\ref{eq:Ecut}), the cutoffs align at the same characteristic energy, where $\kappa=0.65$ has been used for all cases. 

In all simulations, we measured the fraction of dissipated magnetic energy that is converted into ion energy at the time when the ion energy spectrum reaches a steady state. This is reported in Figure \ref{fig3}(c). Across variations in magnetization $\sigma$ and inertial range $l_c/d_i$, approximately $50\%$ of the dissipated magnetic energy is consistently converted into ions, in line with earlier simulations of nonrelativistic turbulence with $\delta B_0 \sim B_0$ \citep{Comisso22}.

\begin{figure}
\begin{center}
    \includegraphics[width=8.65cm]{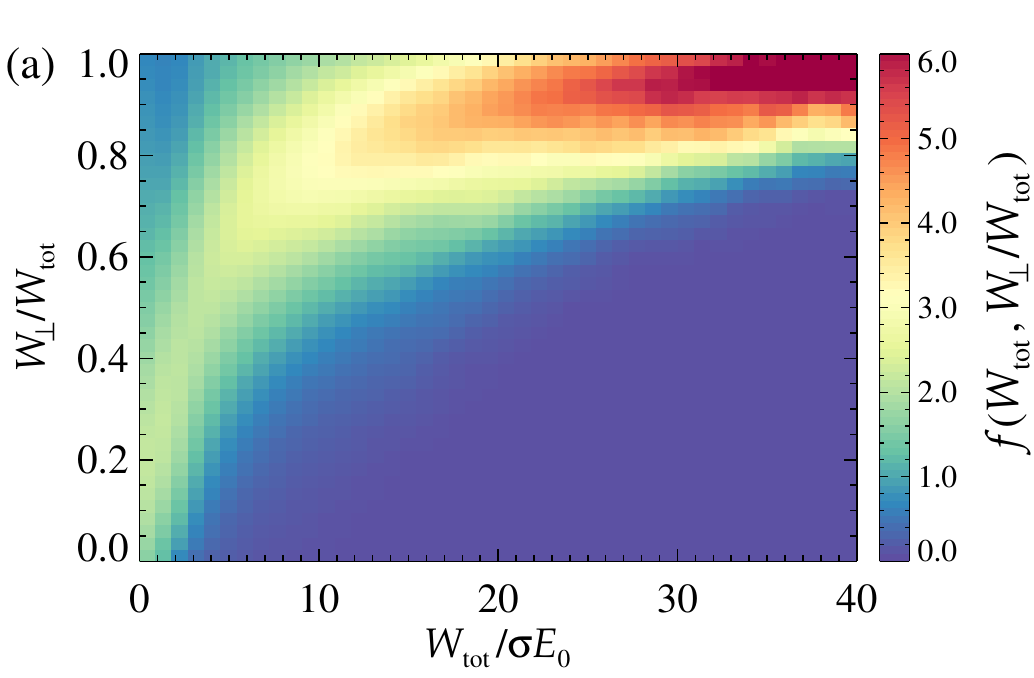}
    
\vspace{0.3cm}

    \includegraphics[width=8.65cm]{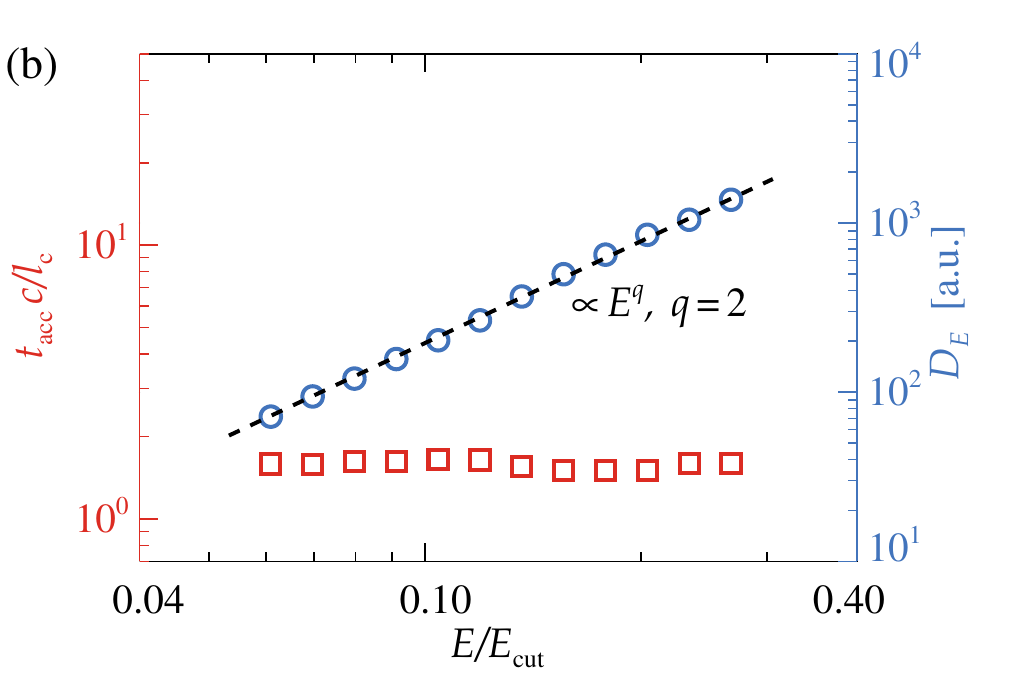}

\vspace{0.12cm}

    \includegraphics[width=8.65cm]{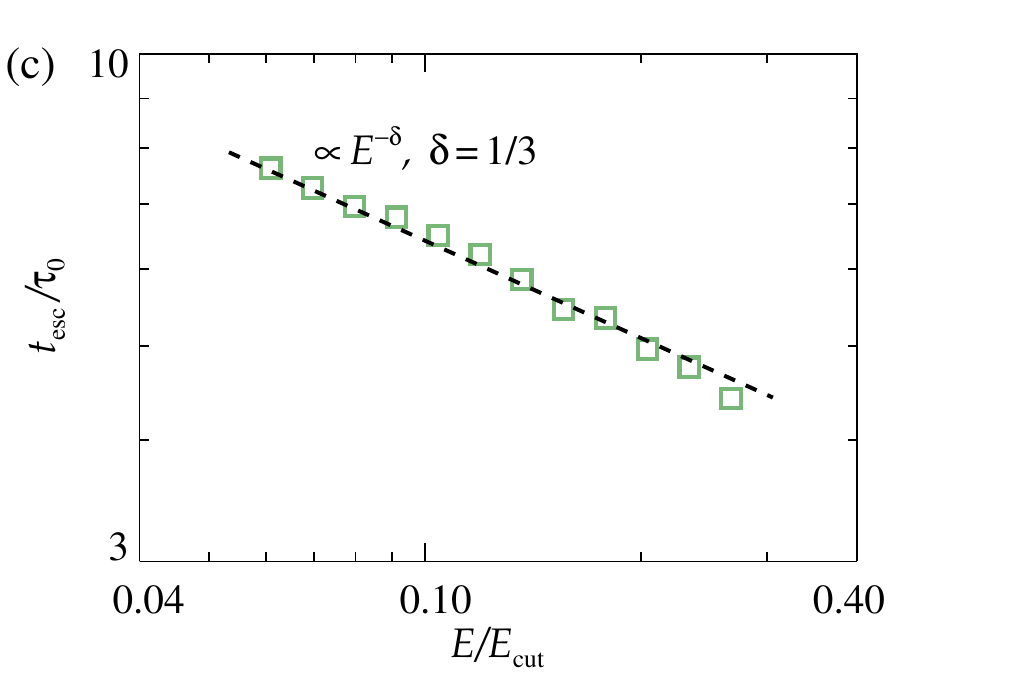}
\end{center}
\vspace{-0.2cm}
    \caption{(a) Particle distribution in terms of $W_{\rm tot}$ and $W_\bot/W_{\rm tot}$ obtained from tracked particles in the reference turbulence simulation ($\sigma = 16$, $l_c/d_i = 133$). (b) Energy diffusion coefficient $D_E$ (blue circles) and stochastic acceleration timescale $t_{\rm acc} = E^2/{4 D_E}$ (red squares), evaluated at $c t_*/l_c = 3$ with a time interval $c \Delta t/l_c = 1.75$. A reference power law $D_E \propto E^{q}$ with a slope of $q = 2$ is also shown. (c) Escape timescale $t_{\rm esc}$, normalized by an arbitrary time $\tau_0$, as a function of particle energy (green squares), evaluated starting at $c t_*/l_c = 3$. A power law $t_{\rm esc} \propto E^{-\delta}$ with slope $\delta = 1/3$ is provided for reference.}
\label{fig4}
\end{figure}

To characterize particle dynamics, we tracked the time evolution of $\sim 3 \times 10^7$ particles (protons) that were randomly selected from our reference simulation. For these particles, we computed the work done by the parallel and perpendicular electric fields, ${W_{\parallel,\bot}}(t) = q \int_0^t {{{\bm{E}}_{\parallel,\bot}}(t') \cdot {\bm{v}}(t') \, dt'}$, where the parallel ($\parallel$) and perpendicular ($\bot$) components are defined with respect to the local magnetic field, i.e. ${\bm{E}}_\parallel = ({\bm{E}} \cdot {\bm{B}}) {\bm{B}}/B^2$ and ${\bm{E}}_\bot = {\bm{E}} - {\bm{E}}_\parallel$. This enables us to isolate the relative contributions of the accelerating electric field components by constructing the distribution
$f ({W_{\rm tot}}, {W_\bot}/{W_{\rm tot}})$, where $W_{\rm tot} = W_\parallel + W_\bot$. This distribution, normalized such that $\int_0^1 {f({W_{\rm tot}},{W_\bot}/{W_{\rm tot}}) \, d( {W_\bot}/{W_{\rm tot}})} = 1$, is shown in Fig. \ref{fig4}(a). When $W_{\rm tot} \lesssim \sigma E_0$, where $E_0$ is the initial particle energy, most of the particle energization occurs via ${\bm{v}} \cdot {\bm{E}}_\parallel$. This energization stage is associated with the non-ideal MHD reconnection electric field, as discussed in \citet{Comisso18,Comisso19}. After this initial phase, particles gain energy via ${\bm{v}} \cdot {\bm{E}}_\bot$ by interacting with large-scale turbulent fluctuations. For particles well into the MHD inertial range of the turbulence cascade, the motional electric field ${\bm{E}}_\bot \simeq - ({\bm{V}}/c) \times {\bm{B}}$ becomes the dominant acceleration channel, with $W_\bot \gg W_\parallel$ for $W_{\rm tot} \gg \sigma E_0$. 

Acceleration via stochastic interactions with turbulent fluctuations can be characterized by an effective diffusion coefficient in energy space, for particles whose gyroradii fall within the turbulence inertial range, $d_i < \rho_L < l_c$. Using the trajectories of particles in the simulation (i.e., not test particles) we compute the mean square energy variation $\langle {{{(\Delta E)}^2}} \rangle$ for particles which have energy $E$ at time $t_*$, based on their energy in the frame comoving with the drift velocity $c{\bm{E}} \times {\bm{B}}/B^2$. The effective diffusion coefficient is then given by $D_E = {\langle {{{(\Delta E)}^2}} \rangle}/{2 \Delta t}$, where $\Delta t  = t - t_*$.  $\Delta t $ must be sufficiently long that the initial conditions have negligible impact and short enough that the turbulence properties have not significantly changed over this interval. For this purpose, we adopt a time interval $c \Delta t/l_c = 1.75$, starting from $c t_*/l_c = 3$. The resulting energy diffusion coefficient as a function of particle energy is shown in Fig. \ref{fig4}(b) with blue circle symbols. We find $D_E \propto E^2$, similarly to what has been found in PIC simulations of electron-positron plasmas \citep{Comisso19, Wong2020} and test-particle simulations in MHD turbulence \citep{Lynn2014,Kimura2018,Sun2021,Pezzi22}. 
Given $D_E$, the stochastic acceleration timescale can be estimated as 
\begin{equation} \label{eq:tacc} 
  t_{\rm acc} = \frac{E^2}{4 D_{E}} ,
\end{equation} 
which is shown in Fig. \ref{fig4}(b) with red square symbols. For the stochastic particle acceleration process relevant here, the acceleration timescale can be expressed as \citep[e.g.][]{Lemoine19,Comisso19}, 
\begin{equation} \label{eq:tacc2} 
  t_{\rm acc} \simeq \frac{1}{4\kappa_{\rm acc} \delta u^2} \frac{B_{\rm rms}^2}{\delta B_{\rm rms}^2} \frac{l_c}{c} 
\end{equation} 
where $\delta u^2 = \Gamma^2 \beta_V^2$ is the squared four-velocity of the scatterers, and $\kappa_{\rm acc} \simeq 0.1$ from PIC simulations \citep{Comisso19}. This stochastic acceleration timescale differs from the commonly assumed, slower, energy-dependent gyroresonant acceleration adopted under the assumption of quasi-linear theory \citep[e.g.][]{Schlickeiser98,Brunetti11}. For large-amplitude turbulence, this stochastic acceleration timescale is comparable to the outer-scale eddy turnover time. 

We next examine the energy dependence of particle escape from the accelerator. The effective scattering mean free path, $\lambda_s$, scales with the Larmor radius as $\lambda_s \simeq l_c \left( \rho_L/l_c \right)^{\delta}$, with $\delta > 0$, namely longer residence times for lower-energy particles. The residence time within the accelerator can then be expressed as 
\begin{equation} \label{eq:tesc} 
  t_{\rm esc} \simeq \frac{L^2}{\lambda_s c} \simeq \frac{L^2}{l_c c} \left( \frac{E_{\rm cut}}{E} \right)^{\delta} \propto E^{-\delta}
\end{equation} 
for ultra-relativistic particles. Here, we evaluate this energy dependence directly from the self-consistent trajectories of the tracked particles.
For particles binned by energy as in the evaluation of $D_E$, we record the time $\Delta t_{l_c}$ it takes for them to move perpendicular to ${\bm{B}}_0$ from their initial coordinates $(x_*,y_*)$ at time $t_*$ and travel a distance greater than $l_c$, i.e. $[(x-x_*)^2+(y-y_*)^2]^{1/2}>l_c$. We consider only particles that remain within the same energy bin and compute the average $\langle{\Delta t_{l_c}}\rangle$ for each energy bin.
The escape time from the accelerator is then given by $t_{\rm esc}(E) \propto \langle \Delta t_{l_c}\rangle(E)$. 
The resulting escape timescale as a function of particle energy is shown in Fig. \ref{fig4}(c) with green square symbols. Our findings indicate that $t_{\rm esc} \propto E^{-1/3}$ offers a reasonable approximation of the PIC simulation results. This scaling is broadly consistent with test-particle simulations \citep[e.g.][]{Casse01,Lemoine23,Kempski23}, which commonly report $\delta$ values in the range $0.3 \leq \delta \leq 0.5$.

The energy-dependent escape timescale results in a hardened escaping particle energy spectrum compared to the spectrum within the accelerator. Accounting for the escape timescale, the flux of particles escaping the accelerator is given by   
\begin{equation} \label{eq:injection} 
  \phi(E) = \frac{dN}{dE dt} = \frac{1}{t_{\rm esc}} \frac{dN}{dE} \propto E^{-s} {\text{sech}}\left[ \left( {\frac{E}{E_{\rm cut}}} \right)^2 \right] \, ,
\end{equation}  
where $s=p-\delta$. The spectral index $s$ characterizes the particles ``injected'' by the accelerator into the surrounding environment. From our turbulence simulations, displaying typical values of $p \sim 2.4$ and $\delta \sim 1/3$, we obtain a characteristic spectral index of $s \sim 2.1$.

\section{UHECR Spectrum and Composition}\label{sec:SpecComp}

\begin{figure*}
    \centering
     \includegraphics[width=17.4cm]{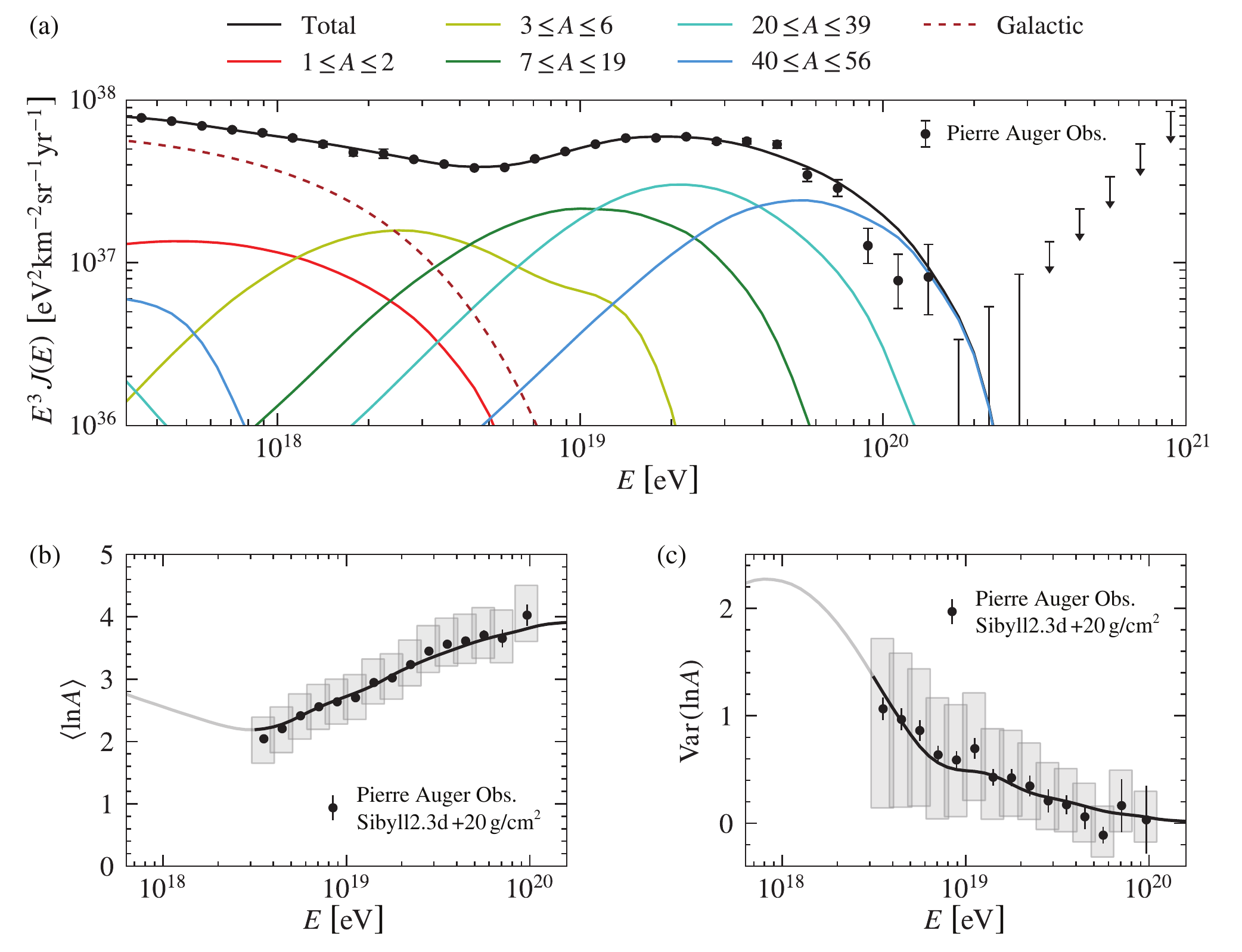}
    \caption{Cosmic ray energy spectrum (top panel) and the first two moments of the $\ln{A}$ distribution (bottom panels) at Earth based on the turbulence acceleration model. In this model, the individual element spectra at the source are given by $\phi_A(E) = f_A \phi_\text{ref} (E/E_\text{ref})^{-s} {\text{sech}}\left[ ( {{E}/{Z e {\mathcal{R}}_{\rm cut}}} )^2 \right]$. The best-fit to the data is given by the spectral index $s = 2.1^{+0.06}_{-0.13}$ and the cutoff rigidity $\log_{10}({\mathcal{R}}_{\rm cut}/\text{V}) = 18.8^{+0.02}_{-0.02}$. Data points are the Pierre Auger Observatory's 2021 spectrum \citep{PierreAuger:2021hun} and composition based on surface detector data \citep{PierreAuger:2024nzw}. 
    Error bars denote the statistical uncertainties (1-sigma statistical errors) and the shaded boxes illustrate the experimental systematic uncertainties (1-sigma systematic errors) of the composition. The composition data is interpreted via the hadronic interaction model \textsc{Sibyll2.3d}~\citep{Riehn:2019jet}.}
    \label{fig5}           
\end{figure*} 

Using the form of the source particle energy spectra given by the turbulence acceleration model, Eq. (\ref{eq:injection}) with Eq. (\ref{eq:Ecut}), and the code developed in~\citep{UFA15,mf23}, we calculate the propagation of the UHECRs starting from their injection into the environment surrounding the accelerator, including their interactions in that region, and continuing with their interactions with the cosmic microwave background (CMB) and the extragalactic background light (EBL) prior to detection.  We explore the multi-dimensional parameter space of injected spectral index, cutoff rigidity, initial composition, as well as key properties of the environment such as peak energy of the photon field and ratio of escape and interaction time, comparing the predicted spectrum and composition observables to the spectrum and nuclear composition observed at Earth. 
The treatment takes into account the following interactions with the ambient photon field: (i) pion photoproduction $p + \gamma \rightarrow \pi + p$ (which, for interactions with CMB photons, leads to the so called Greisen-Zatsepin-Kuzmin cutoff \citep{Greisen66,ZK66}), (ii) photo-disintegration of nuclei in both the source environment and during extragalactic propagation via interactions with CMB and EBL photons, $A + \gamma \rightarrow (A-nN)+nN$, where $n$ is the number of nucleons lost by the nucleus, (iii) pair photoproduction $p + \gamma \rightarrow e^+ + e^- + p$ (Bethe-Heitler process) during extragalactic propagation. Extragalactic propagation was simulated using \textsc{CRPropa3}~\citep{AlvesBatista:2016vpy}, adopting the EBL model of~\citet{Gilmore:2011ks} and a star formation rate (SFR) source evolution~\citep{Robertson:2015uda}. 

A combination of different injection elements is used as an approximation of different mass groups in the cosmic-ray composition. Each of these elements is characterized by the same power-law index $s$ and their spectra are given by $\phi_A(E) = f_A \phi_\text{ref} (E/E_\text{ref})^{-s} {\text{sech}}\left[ ( {{E}/{E_{\rm cut}}} )^2 \right]$, where $E_{\rm ref} = 10^{17}$ eV is a reference energy selected below the minimum cutoff energy for protons, $\phi_\text{ref}$ is a normalization constant, and the injection fractions $f_A$, which are defined as relative number flux ratios at the same reference energy $E_{\rm ref}$, are determined through the fit. We approximate the Galactic cosmic ray component as a single $A=38$ composition, with the spectral index and composition treated as auxiliary parameters. For details see~\cite{mf23,mfu22,UFA15}.

We fit the model predictions to the UHECR spectrum and composition data of the Pierre Auger Observatory \citep{Aab+20a,Aab+20b,PierreAuger:2021hun,PierreAuger:2024nzw} as detailed in~\cite{mf23}. We map $\langle X_\mathrm{max} \rangle$ and $\mathrm{Var}\left(X_\mathrm{max}\right)$ data into $\langle \ln{A} \rangle$ and $\mathrm{Var}(\ln{A})$ using the parametrization of the \cite{Abreu+13}, after shifting the $\langle X_\mathrm{max} \rangle$ predictions of the hadronic interaction model \textsc{Sibyll2.3d}~\citep{Riehn:2019jet} by $+20$~g/cm$^2$ following the~\cite{PierreAuger:2024neu}. To assess the goodness-of-fit we compute a combined $\chi^2$ to the spectrum data and composition,
\begin{align}
    \chi^2 = &\displaystyle\sum_i^{N_\mathrm{spec}} \frac{(J_{m,i}-J_i)^2}{\sigma_{J,i}^2} + \displaystyle\sum_j^{N_\mathrm{comp}} \frac{(\langle \ln{A} \rangle_{m,j} - \langle \ln{A} \rangle_j )^2}{\sigma_{\langle \ln{A} \rangle,j}^2} \nonumber \\
    &+ \displaystyle\sum_j^{N_\mathrm{comp}} \frac{(\mathrm{Var}(\ln{A})_{m,j} - \mathrm{Var}(\ln{A})_j)^2}{\sigma_{\mathrm{Var}(\ln{A}),j}^2},
\end{align}
where $N_\mathrm{spec}$ and $N_\mathrm{comp}$ are the number of data points in the spectrum and composition, respectively, ${\mathcal{Q}}_{m,i}$ denotes the model prediction for the quantity $\mathcal{Q}$ (differential flux $J$, mean logarithmic mass number $\langle\ln{A}\rangle$, or its variance $\mathrm{Var}(\ln{A})$) at energy bin $i$, and errors include only statistical errors for the energy spectrum from the~\cite{Aab+20a,Aab+20b,PierreAuger:2021hun} and the mean and variance of $\ln{A}$ from the~\cite{PierreAuger:2024nzw}.  For spectral energy bins above the highest-energy data point, we follow \cite{Baker84} by adding an additional $2n_i$ to the $\chi^2$, where $n_i$ is the expected number of observed events predicted by the model in energy bin $i$ given the exposure of the dataset. Our final figure of merit is $\chi^2_\mathrm{tot} = \chi^2 + 2\sum_i n_i$, where $i$ runs over energy bins above the highest-energy data point in the spectrum.

\begin{figure}
\begin{center}
    \includegraphics[width=8.00cm]{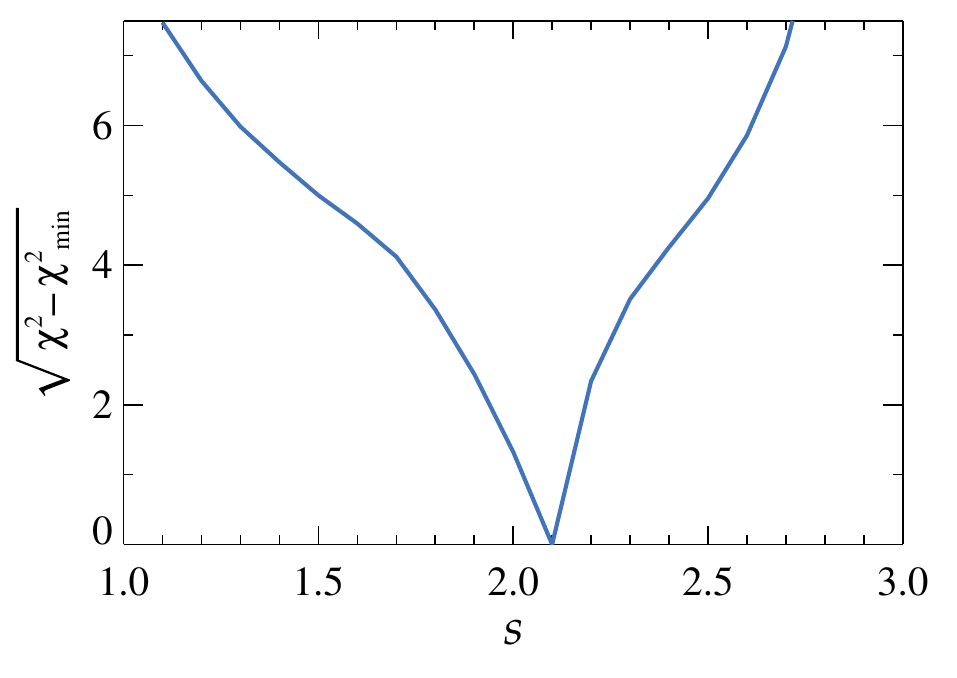}

\end{center}
\vspace{-0.5cm}
    \caption{$\sqrt{\chi^2-\chi^2_{\rm min}}$ as a function of the source spectral index $s$ for the turbulence acceleration model, where the individual element spectra from the accelerator are given by Eq.~(\ref{eq:injection}) with $E_{\rm cut} = Z e {\mathcal{R}}_{\rm cut}$.}
    \label{fig6}   
\end{figure}

Figure \ref{fig5} shows the predicted energy spectrum $J(E)=dN/dEdtdAd\Omega$, multiplied by $E^3$, along with the composition moments $\langle\ln{A}\rangle$ and $\mathrm{Var}(\ln{A})$, based on the best-fit model parameters. 
We explored source spectral indices within the range $1<s<3$, motivated by standard electromagnetic acceleration mechanisms. The turbulence-based model results in a best-fit source spectral index of $s = 2.1^{+0.06}_{-0.13}$, based on the $\chi^2$ goodness-of-fit test (see Fig.~\ref{fig6}), with a cutoff rigidity $\log_{10}({\mathcal{R}}_{\rm cut}/\text{V}) = 18.8^{+0.02}_{-0.02}$. We note that the exact best-fit value of the spectral index and its uncertainty are sensitive to the composition dataset used for the fitting, but are qualitatively unchanged. 

As could be expected, the turbulence-based ${\text{sech}}\left[ ( {{E}/{E_{\rm cut}}} )^2 \right]$ cutoff compresses the energy range over which each element group significantly contributes, as required for consistency with the narrowness of the $\ln{A}$ distribution. This results in a relatively soft power-law index at the source ($s \sim 2 - 2.2$) that is consistent with our PIC simulation results.  By contrast, the same analysis but using the exponential cutoff gives a worse fit and requires a spectral index $s = 1.26^{+0.1}_{-0.1}$, incompatible with diffusive shock acceleration theory. It is worth noting that in both cases, the power-law spectrum leaving the source environment is roughly one unit harder than that given by the accelerator, due to lower-rigidity particles undergoing more interactions before escaping the source environment \citep{UFA15}.

For the best-fit results shown in Fig.~\ref{fig5}, taking into account all element groups, we obtain that the volumetric energy-injection rate of UHECRs, $\mathcal{L}_{_{\rm UHECR}}$, defined as the total energy injected per unit of comoving volume per unit time at redshift~$z=0$, is 
\begin{equation} \label{eq:L} 
\mathcal{L}_{_{\rm UHECR}} \! = \! \sum_A \! \int_{E_a}^{E_b} \! \! E  {Q}_A(E) dE \simeq 0.7 \times 10^{45}\text{erg\,Mpc}^{-3} \text{yr}^{-1} ,
\end{equation} 
where ${Q}_A(E)$ indicates the number of nuclei with atomic mass number $A$ injected per unit of energy, volume and time, and the integration limits are $E_a = 10^{18}$ eV and $E_b = 10^{21}$ eV. This volumetric energy-injection rate, which is about a factor-two larger than found with the exponential cutoff with its associated required harder spectral index, is compatible with different source classes, including jets from AGNs, gamma-ray bursts, and tidal disruption events, where magnetized turbulence is expected to be at play.

\section{Conclusions}\label{sec:conclusions} 

We have demonstrated, using fully kinetic particle-in-cell simulations, that particle acceleration by magnetically dominated turbulence possesses the properties needed to explain the acceleration of UHECRs. After a low-energy injection phase driven by magnetic-field-aligned electric fields, ions are accelerated by scattering off turbulent fluctuations, gaining energy through the motional electric field, up to the cutoff energy $E_{\rm cut} = Z e {\mathcal{R}}_{\rm cut} = Z e B_{\rm rms} \kappa l_c $, with $\kappa \simeq 0.65$. This acceleration process occurs on a short, energy-independent timescale, which, for large amplitude turbulence, is comparable to the outer-scale eddy turnover time. The resulting particle energy spectrum follows a power law with a sharp cutoff at the characteristic energy scale $E_{\rm cut}$. Taking into account the energy dependence of particle escape from the accelerator, the spectrum of relativistic ions produced by magnetically dominated turbulence at the source is well-approximated by ${dN}/{dE dt} \propto E^{-s} {\text{sech}}\left[ ( {{E}/{E_{\rm cut}}} )^2 \right]$ with $s \sim 2.1$. The ${\text{sech}}\left[ ( {{E}/{E_{\rm cut}}} )^2 \right]$ cutoff is considerably sharper than the $\exp \left(-{E}/{E_{\rm cut}} \right)$ cutoff expected for diffusive shock acceleration, enabling softer particle spectra at the source to explain the UHECR spectrum and composition observed at Earth.

We have fit the Pierre Auger Observatory's spectrum and composition measurements, accounting for particle interactions between acceleration and detection, and found that the turbulence-associated ${\text{sech}}\left[ ( {{E}/{E_{\rm cut}}} )^2 \right]$ cutoff with a spectral index of $s \sim 2 - 2.2$ is supported by the data. This result contrasts with the significantly harder power-law indices needed by exponential cutoff models inspired by diffusive shock acceleration \citep{Aab2017JCAP,Halim2023JCAP}, which are incompatible with the expected $s \gtrsim 2$. Other acceleration mechanisms, such as magnetic reconnection, could potentially produce harder ion energy spectra with $s \sim 0-2$ \citep{Comisso24}, but only over a limited energy range $E \lesssim \sigma_{\delta B} m_i c^2$. For energies $E \gg \sigma_{\delta B} m_i c^2$, the asymptotic power-law index from magnetic reconnection returns to $s \sim 2$. Another potential way for producing a sharper-than-exponential cutoff involves energy losses in the accelerator, such as synchrotron losses, which would result in a cutoff of the form $\exp \left[-({{E}/{E_{\rm cut}}})^2\right]$ \citep{zirakashviliAharonian06}. However, the need for finely tuned energy losses and the absence of pile-up in the observed spectrum of UHECRs~\citep{Protheroe04,zirakashviliAharonian06} strongly disfavors this scenario. 

Analytic analysis~\citep{ProtheroeStanev99} suggests that the spectral cut-off in diffusive shock acceleration is exponential or softer, with spectral index $s \gtrsim 2$.  If this behavior is born out by the analysis of PIC simulations for shock acceleration, as we have done for magnetized turbulence, it will be possible to use the combined fit of UHECR spectrum and composition data to discriminate between acceleration mechanisms.  While a comprehensive comparison of fits with different cutoffs is left for future work, within the framework of the analysis reported in Sec.~\ref{sec:SpecComp}, the exponential cutoff both gives a worse fit and requires a harder spectral index  incompatible with theoretical expectation.  

Highly magnetized turbulence, as considered in this study, is expected in a range of astrophysical environments, such as jets from active galactic nuclei \citep{Blandford19}, outflows from neutron star mergers \citep{Kiuchi24}, jets from collapsars \citep{Thompson94}, jetted tidal disruption events \citep{Curd19}, and coronae of Seyfert galaxies \citep{Murase20}. Determining which of these systems can meet the specific conditions required for the particle acceleration mechanism discussed here will be an essential next step in identifying the sources of UHECRs.

\vspace{0.5cm}

\section*{Acknowledgments}
We thank Michael Unger and Teresa Bister for their feedback on the manuscript and Joseph Insley for his assistance with the visualizations.
L.C. was supported by the National Science Foundation award PHY-2308944, G.R.F. by NSF PHY-2013199, and M.S.M. by the NSF MPS-Ascend Postdoctoral Award \#2138121. Computational resources were provided by the Innovative and Novel Computational Impact on Theory and Experiment (INCITE) program, using resources of the Argonne Leadership Computing Facility, which is a DOE Office of Science User Facility supported under Contract No. DE-AC02-06CH11357. Additional computational support was provided through Columbia University’s Shared Research Computing Facility and NYU IT High Performance Computing resources, services, and staff expertise. We acknowledge the use of code created by Michael Unger under the EU-funded Grant PIOF-GA-2013-624803 and further developed by Marco Muzio.

\newpage

\bibliography{turb_acceleration}{}
\bibliographystyle{aasjournal}

\end{document}